\documentclass[eqsecnum,aps,superscriptaddress,11pt ]{revtex4}
\usepackage{graphicx}
\usepackage{dcolumn}
\usepackage{xspace}

\begin{document}
\title{Many-Body Effects in Hyperfine Interactions in $^{205}$Pb$^+$}
\vspace*{0.5cm}

\author{Sonjoy Majumder$^1$, \protect\footnote[3]{Electronic address: bijaya@mpipks-dresden.mpg.de} B. K. Sahoo$^{2}$, R. K. Chaudhuri$^3$, B. P. Das$^3$ and D. Mukherjee$^4$ \\
\vspace{0.3cm}
{\it $^1$Department of Physics, Indian Institute of Technology-Madras, Chennai-600 036, India} \\
{\it $^2$Max Planck Insttitute for the Physics of Complex Systems, D-01187 Dresden, Germany} \\
{\it $^3$Non-Accelerator Particle Physics Group,
Indian Institute of Astrophysics, Bangalore-34, India}\\
{\it $^4$Department of Physical Chemistry,\\ 
Indian Association for Cultivation of Science, Calcutta-700 032, India} }
\date{\today}

\begin{abstract}
\noindent
{\it Ab initio} calculations have been carried out to study the magnetic dipole and 
electric quadrupole hyperfine structure constants of $^{205}$Pb$^+$. Many-body 
effects have been considered to all orders using the relativistic coupled-cluster theory in the singles, doubles and partial triples approximation. The trends of these 
effects are found to be different from atomic systems that have been studied 
earlier. 
\end{abstract} 
\maketitle

\section{Introduction}
With the advent of ion trapping techniques, it has become possible to perform  
high precision measurements of different atomic properties; including hyperfine structure constants
for the ground and low-lying excited states of stable isotopes \cite{werth,liu}.
Studies of these interactions have served as stringent tests of relativistic
many-body theories. It has been found that the behavior of electron correlation in the hyperfine interactions in the $d$-states is substantially different 
from those of $s$- and $p$- states \cite{sahoo3}. The underlying reason for this
is the strong and unusual core-polarization effects associated  
with the former.

The work presented in this paper is carried out using the relativistic coupled-cluster (RCC) theory. This theory has
certain features which make it suitable for the calculations of excited state 
properties. Prominent among them being its abilities to treat relativistic and correlation 
effects in the initial and final states in a rigorous and balanced way \cite{sahoo2}. It
is equivalent to all orders relativistic many-body perturbation theory. Core-polarization
and pair-correlation effects which are important for such calculations are computed to
all orders in this theory. It has been successfully applied to heavy atomic systems with
a single valence electron \cite{geetha,sahoo2}.
  
$^{205}$Pb is one of the stable isotopes of lead  which has lifetime about 1.5 $\times \ 10^7$ years. 
The nuclear structure of this isotope is different from the other stable isotope, $^{207}$Pb. 
It's nuclear magnetic moment and electric quadrupole moment are non-zero.
Theoretical studies of the magnetic dipole and electric quadrupole hyperfine structure constants of $^{205}$Pb$^+$
are the focus of the present work. 

There are relatively few measurements of the hyperfine structure constants for  the excited states of heavy atoms and ions.
It is certainly worthwhile to perform highly correlated calculations of these  
quantities for Pb$^+$ as it could be a promising candidate for the observation of parity non-conservation \cite{bijaya}. This theoretical effort could motivate
experimentalists to carry out high precision measurements of $^{205}$Pb$^+$.

In section II of this paper, we give an outline of the RCC theory and in section
III we discuss the pertinent details of the calculation. The results of our
calculations are presented and discussed in section III and finally we make some
concluding remarks in section IV.

\section{Theory}
In order to obtain the RCC wavefunction for Pb$^+$, we require the 
closed-shell Pb$^{++}$ wavefunction, which is given in terms of 
the lowest order Dirac-Fock (DF) wavefunction 
$|\Phi_{DF}\rangle$, as
\begin{equation}
|\Psi\rangle=\Omega|\Phi_{DF}\rangle,
\end{equation}
where in conventional many-body perturbation theory, the wave operator, 
$\Omega$ is expressed 
in powers of the residual interaction, 
$V_{es} = \sum_{i<j} \frac {1} {r_{ij}} - \sum_{i} U_{DF}(r_i)$.
This results in a hierarchy of approximations for the correlation energy and 
the wavefunction.

The coupled cluster theory is based on the following exponential ansatz for 
the closed-shell wave operator \cite{lindgren}
\begin{equation}
\Omega = N [exp(T)],
\end{equation}
where the cluster operator $T$ is expressed in terms of the connected 
diagrams of the wave operator.
The operator $T$ also can be expressed in terms of the cluster operators 
$T_n$ corresponding to different orders of excitations 
$n$ of the core electrons from the 
DF state,
$|\Phi_{DF}\rangle$, explicitly defined by,
\begin{eqnarray}
T &=& T_1 + T_2 + ...\nonumber \\
&=& \sum_{ap}a_p^{\dag}a_a t_a^p + \frac {1}{4}\sum_{abpq}a_p^+a_q^+ 
a_ba_a t^{pq}_{ab} + ... ,
\end{eqnarray}
with $a,b,c,.. (p,q,r,..)$ representing occupied (unoccupied) orbitals. 
$t_a^p$ is the cluster amplitude corresponding to the single excitation from the
orbital $a$ to $p$ and so on. Termination of the series at 
$T_2$, results in the coupled cluster theory with single and double excitations
(CCSD). The contribution from the Breit interaction \cite{breit} which is four orders of magnitude
smaller than the Coulomb interaction has been neglected in the present work.   

For a single valence atomic system the wavefunction in the RCC method can be written 
as \cite{lindgren,debasish}
\begin{equation}
|\Psi_v\rangle = e^T\{1+S_v\}|\Phi_v\rangle,
\end{equation}
with the new reference state 
\begin{equation}
|\Phi_v\rangle = a_v^{\dagger} |\Phi_{DF}\rangle,
\end{equation}
for the given valence electron $v$;  $S_v$ represents excitation operators 
which excite at least  
the valence electron. The explicit form of this operator can be written as  
\begin{eqnarray}
S_v &=& S_{1v} + S_{2v} + ... \nonumber \\
&=& \sum_{p \ne v}a_p^+a_v s_v^p + \frac {1}{2}\sum_{bpq}a_p^+a_q^+a_ba_v s^{pq}_{vb} + ...,
\end{eqnarray}

An approximate treatment of the triple excitations to the CCSD method is included by contracting
the residual Coulomb operator,  which effectively forms a
two-body operator, and the double-excitation operators $T_2$
and $S_{2v}$ \cite{haque}, thereby defining the CCSD(T) approximation
\begin{equation}
S_{vbc}^{pqr}\ =\ \frac{\widehat{V_{es}T_2}+\widehat{V_{es}S_{2v}}}{\epsilon_v+\epsilon_b+\epsilon_c-\epsilon_p-\epsilon_q-\epsilon_r},
\end{equation}
where $\epsilon_i$ is the orbital energy of the corresponding {\it i'th} electron.

\section{Method of Calculation}
For computational simplicity, the $T$ amplitudes are solved 
first for the closed-shell Pb$^{++}$ and then the valence electron 
is attached to calculate the open-shell wavefunctions. The matrix equations for solving the correlation energy and 
the $T$ amplitudes are given by
\begin{eqnarray}
\langle \Phi_{DF}|\overline{\text{H}_N} |\Phi_{DF}\rangle = \Delta E_{corr}  \\
\langle \Phi_{DF}^* |\overline{\text{H}_N} |\Phi_{DF} \rangle = 0,
\end{eqnarray}
where $\text{H}_N$ is the normal ordering Hamiltonian which can be written as 
$\text{H}_N=\text{f}_N + \text{V}_N$, sum of one-body and two-body terms. In the above equation, 
$\overline{\text{H}_N}$ is defined as 
$\overline{\text{H}_N} = e^{-T} \text{H}_N e^T$, $\Delta E_{corr}$ 
is the correlation energy and $|\Phi_{DF}^*\rangle$ corresponds to excited states
from $|\Phi_{DF} \rangle$.

For the open-shell RCC calculations, the $S_v$ operators are solved using the following 
equations
\begin{eqnarray}
\langle \Phi_v|\overline{\text{H}_N} \{1+S_v\}|\Phi_v\rangle = -\Delta E_v  \\
\langle \Phi_v^{*}|\overline{\text{H}_N} \{1+S_v\}|\Phi_v\rangle = -\Delta E_v \langle \Phi_v^{*}|
\{S_v\}|\Phi_v\rangle,
\end{eqnarray}
where $\Delta E_v$ is the ionization potential (IP) energy of the corresponding valence 
electron $v$. Similarly $|\Phi_v^{*}\rangle$ are the excited states from $|\Phi_v\rangle$.

The relativistic hyperfine Hamiltonian is given by \cite{cheng}
\begin{equation}
H_{hfs} = \sum_k \bf{M}^{(k)}\cdot \bf{T}^{(k)},
\end{equation}
where $\bf{M}^{(k)}$ and $\bf{T}^{(k)}$ are spherical tensor operators of
rank k. In first-order perturbation theory, hyperfine energies $E_{hfs}(J)$ of the 
fine-structure state $|JM_J\rangle$ are expectation values of the hyperfine interaction Hamiltonian.
Details of the expression are given by Chang and Childs \cite{cheng}. The
magnetic  dipole and electric quadrupole hyperfine energies are defined by
\begin{equation}
E_{M1}=AK/2
\end{equation}
and 
\begin{equation}
E_{Q2}={B\over 2}\frac {3K(K+1)-4I(I+1)J(J+1)}{2I(2I-1)2J(2J-1)},
\end{equation}
respectively. Here $I$ and $J$ being the total angular momentum of the nucleus and the electron
state, respectively, and $K=2\langle I\cdot J\rangle$. The magnetic dipole hyperfine
constant $A$ and electric quadrupole hyperfine constant $B$ are defined as 
\begin{equation}
A=\mu_Ng_I\frac{\langle J ||T^{(1)}||J\rangle}{\sqrt{J(J+1)(J+1)}}
\end{equation}
and
\begin{equation}
B=2eQ\left[\frac{2J(2J-1)}{(2J+1)(2J+2)(2J+3)}\right]^{1/2}\langle J ||T^{(2)}||J\rangle ,
\end{equation}
respectively, where $\mu_N$ is Bohr magneton, $g_I=\mu_I/I$ with $\mu_I$ and $I$ are the nuclear dipole moment and spin,
and $Q$ is nuclear quadrupole moment.

The expectation value for a general one particle operator $O$ in a given valence electron ($v$) state can be expressed in RCC theory as
\begin{eqnarray}
 \langle O \rangle_v  &=& \frac {\langle\Psi_v | O | \Psi_v \rangle} {\langle\Psi_v|\Psi_v\rangle} \nonumber \\
 &=& \frac {\langle \Phi_v | \{1+S_v^{\dagger}\} e^{T^{\dagger}} O e^T \{1 + S_v\} | \Phi_v\rangle } {1+N_v} \nonumber \\
 &=& \frac {\langle \Phi_v | \{1+S_v^{\dagger}\} \overline{O} \{1 + S_v\} | \Phi_v\rangle } {1+N_v},
\end{eqnarray}
where we define
\begin{eqnarray}
\overline{O}= e^{T^{\dagger}} O e^T
\end{eqnarray}
and
\begin{eqnarray}
N_v &=& \langle \Phi_v | S_v^{\dagger} [e^{T^{\dagger}} e^T] + S_v^{\dagger} [e^{T^{\dagger}} e^T] S_v + [e^{T^{\dagger}} e^T] S_v | \Phi_v\rangle\nonumber \\
 &=& \langle \Phi_v | S_v^{\dagger} \overline{n_v} + S_v^{\dagger} \overline{n_v} S_v + \overline{n_v} S_v^{\dagger} | \Phi_v\rangle .
\end{eqnarray}

For computational simplicity we evaluate the matrix elements of any operator
in two steps. 
We expand $\overline{O}$ using Wick's general theorem \cite{lindgren} as
\begin{eqnarray}
\overline{O} &=& (e^{T^{\dagger}} O e^T)_{f.c.} + (e^{T^{\dagger}} O e^T)_{o.b.} + 
(e^{T^{\dagger}} O e^T)_{t.b.} + ...., 
\end{eqnarray}
where we have used the abbreviations f.c., o.b. and t.b. for fully
 contracted, effective 
one-body and effective two-body terms respectively. In this expansion of $\overline{O}$, 
the effective one-body terms are computed keeping terms of the form of
\begin{eqnarray}
\overline{O}_{o.b.} &=& O + T^{\dagger} O + O T + T^{\dagger} O T .
\end{eqnarray}

The calculation procedure for these terms are given by Geetha {\it et al} \cite{geetha}.
They are finally connected with $S_v$ and $S_v^{\dagger}$ operators in the
evaluation of properties . Contributions due to the effective two-body terms from 
$\overline{O}$ are constructed using the procedure shown diagrammatically in
our earlier works \cite{geetha,bijaya1} and computed directly during the
calculation of properties.The following 
The following types of terms are considered for the construction of the effective two-body terms
\begin{eqnarray}
\overline{O}_{t.b.} = O T_1 + T_1^{\dagger} O + O T_2 + T_2^{\dagger} O.
\end{eqnarray}
Other effective terms correspond to higher
orders in the residual Coulomb interaction and are neglected in the
present calculation.
A similar procedure has been followed to account for the normalization factor.

The contributions from the normalization factors for the corresponding valence electron $v$ that are given in
in tables III and IV, were obtained using the following 
relations:
\begin{eqnarray}
Norm = \langle \Psi_v | O | \Psi_v \rangle \{ \frac {1}{1+N_v} - 1 \}.
\end{eqnarray}

\section{Results and Discussions}
The starting point of our calculations is the generation of DF orbitals for $^{205}$Pb$^{++}$. These orbitals are   
constructed as linear combinations of Gaussian type orbitals (GTOs) as given
by Chaudhuri et al. \cite{rajat}. It has been found \cite{bijaya} that
the RCC  calculations  based on these  kind of GTOs provide accurate $A-$ values
for the low-lying states of $^{207}$Pb$^+$. In the present work, we have used
similar basis functions to calculate both 
$A-$ and $B-$ values of ground and some of the important excited states of $^{205}$Pb$^+$.

In table I, we present A and B hyperfine structure constants
of the low-lying states of Pb$^+$. We use the Lande nuclear 
g-factor, $g_I$ = 0.28468 to calculate  $A$ and nuclear quadrupole 
moment, $Q$ = 0.234 to calculate B \cite{Anselment}.  From the differences of the DF and RCC 
results given in table I, it is evident that for both the hyperfine constants A and B the electron 
correlation effects vary from (10-290)\%. 
  
\begin{table}[h]
\caption{$A$ and $B$ results of $^{205}$Pb$^+$ using DF and RCC methods.}
\begin{ruledtabular}
\begin{center}  
\begin{tabular}{lcccccccc}
& $6p_{1/2}$ & $6p_{3/2}$ & $7s_{1/2}$ &  $7p_{1/2}$ & $7p_{3/2}$ & $6d_{3/2}$ & $6d_{5/2}$ & $8s_{1/2}$ \\
\hline\\
\underline{$A$} \\
DF & 2765.54 & 220.58 & 1879.06 & 476.34 & 43.84 & 19.05 & 7.53 & 687.77 \\
RCC & 3099.5 & 149.7 & 2680.3 & 543.7 & 74.8 & -10.03 & 60.97 & 929.3 \\ 
\hline\\
\underline{$B$} \\
DF & & 377.81 & & & 75.09 & 11.08 & 13.07 &  \\ 
RCC & & 464.6 & & & 99.3 & 50.7 & 56.9 &  \\ 
\end{tabular}        
\end{center}
\end{ruledtabular}
\label{tab1}
\end{table}
All the core orbitals were excited in our calculations. The core-polarization 
effects, which are the largest contributors to the hyperfine constants of the $6p_{3/2}$ state of 
$^{207}$Pb$^+$ \cite{bijaya} and the $d_{5/2}$ states of the alkaline earth ions \cite{bijaya2}, have been 
accounted to all orders through the $OS_{2v}$ term.  It was also found from 
the hyperfine structure studies of the $s_{1/2}$ and $p_{1/2}$ states in the alkaline earth 
ions that pair-correlation and core-polarization effects are important. 
In order to appreciate the importance of these effects in Pb$^+$, we present their 
contributions in table II. Comparison of these results 
with their corresponding DF values from Table I, brings out some distinct many-body features 
of the system. The most prominent among them is the size of the core-correlation and core-polarization 
effects for the $d-$ states. As in the case of some of the alkaline earth ions, the sign of the core-polarization effect 
in the $d_{5/2}$ state is opposite that of the DF value and the net contribution
is 209\% of the corresponding DF value. However, unlike those alkaline earth ions \cite{bijaya2} the final 
RCC result has the same sign as the DF result. This is due to the large positive contributions from the 
$S_{2v}^{\dagger}\overline{O}S_{2v}$ term, shown in table III . Therefore, it is imperative to consider such terms to obtain accurate results. 

The role of electron correlation in the hyperfine interactions in $^{45}Sc$ and $^{89}Y$ has similarities
\cite{sahoo3} with Pb$^+$ even though those systems are neutral and have different
electronic configurations. Another
interesting feature of the present 
study on $^{205}$Pb$^+$ is that the RCC result of the $A-$ value of $6d_{3/2}$ is of opposite in sign than that of
the DF result. This trend is different from the other $d_{3/2}$ states in alkaline earth ions \cite{sahoo3,csur}. 
The main reason for this behavior is due to another unusual contribution from 
$S_{2v}^{\dagger}\overline{O}S_{2v}$. The core correlation effect on $B-$ values though reduces as higher 
excited states are considered, which is expected, but contributions to the $d-$ states are not 
similar to the $p-$ states. This is evident from the $B-$ values given in table II. 

Table III and IV present the important effective two-body terms obtained from  
$\overline{O}= e^{T^{\dagger}}Oe^T$, but they contribute very little. One can
therefore justifiably ignore the higher order terms 
given in eqn. (3.15) and save computational time. The 
correlation effects of the $A-$ values of other states presented in table I, behave
the same way as in $^{207}$Pb$^+$  \cite{bijaya}.

\begin{table}
\caption{The contributions of core-correlation, core-polarization and pair-correlation of
$A$ and $B$ results in $^{205}$Pb$^+$.}
\begin{ruledtabular}
\begin{tabular}{lrrr}
States          &  Core-corr. & Core-pol.    &  Pair-corr.  \\
\hline\\
A\\
$7p_{3/2}$   &  -0.68     & 15.17    &  6.01   \\ 
$6d_{3/2}$   &   -1.02    & 11.13    &  9.30   \\
$6d_{5/2}$   &   -0.48    & -15.73    &  3.45   \\
$8s$         &   70.29   & 144.23   & 139.92   \\
\hline\\
B\\
$6p_{3/2}$ &   29.26   &  79.19   &  35.55    \\
$7p_{3/2}$ &   3.34    & 15.78    & 10.29   \\
$6d_{3/2}$ &  -0.65    & 17.87    &  5.44   \\
$6d_{5/2}$ &  -0.88    & 23.46    &  5.96   \\
\end{tabular}
\end{ruledtabular}
\label{tab:results1}
\end{table}

\begin{table}[h]
\caption{Contributions of different coupled-cluster terms to the Pb$^+$
 magnetic dipole hyperfine structure constant ($A$).  $cc$ stands for the complex conjugate part of the corresponding terms.}
\label{tab:front3}
\begin{tabular}{lcccc}
\hline
\hline
Terms & 7p$_{3/2}$ & 6d$_{3/2}$ & 6d$_{5/2}$ & 8s$_{1/2}$ \\
 & state & state & state & state \\
\hline
 & & & & \\
  $O$ (DF) & 43.84 & 19.05 & 7.53 & 687.77 \\
\hline
$\overline{O}$ & 44.52 & 20.07 & 8.01 & 617.48 \\
$\overline{O} S_{1v} + cc $ & 6.01 & 9.30 & 3.45 & 139.92 \\
$\overline{O} S_{2v} + cc $ & 15.17 & 11.13 & -15.73 & 144.23 \\
$S_{1v}^{\dagger} \overline{O} S_{1v}$ & 0.21  & 1.09 & 0.38 & 7.92 \\
$S_{1v}^{\dagger} \overline{O} S_{2v} + cc $ & -0.86 & 2.02  & -2.08 & 4.85 \\
$S_{2v}^{\dagger} \overline{O} S_{2v} + cc $ & 11.04 & -54.93 & 68.91 & 29.74 \\
\hline\\ 
\multicolumn{5}{c}{\textbf{Important effective two-body terms of $\overline{O}$ }} \\
\hline
 & & & & \\
$S_{2v}^{\dagger} O T_1 + cc $ & -0.03 & -0.03 & -0.01 & 1.03 \\
$S_{2v}^{\dagger} O T_2 + cc $ & -0.44 & 0.74 & 0.26 & -3.61 \\
 $Norm$ & -0.69 & 0.39 & -2.21 & -10.06 \\
\hline
\hline
\end{tabular}
\end{table}

\begin{table}[h]
\caption{Contributions of different coupled-cluster terms to the Pb$^+$
electric quadrupole hyperfine structure constant ($B$).}
\begin{tabular}{lcccc}
\hline
\hline
Terms & 6p$_{3/2}$ & 7p$_{3/2}$ & 6d$_{3/2}$ & 6d$_{5/2}$ \\
 & state & state & state & state \\
\hline
 & & & & \\
 $O$ (DF) & 377.81 & 75.09 & 11.08 & 13.07 \\
\hline
$\overline{O}$ & 348.55 & 71.56 & 11.73 & 13.95 \\
$\overline{O} S_{1v} + cc $ & 35.55 & 10.29 & 5.44 & 5.96 \\
$\overline{O} S_{2v} + cc $ & 79.19 & 15.78 & 17.87 & 23.46 \\
$S_{1v}^{\dagger} \overline{O} S_{1v}$ & 0.87  & 0.36 & 0.64 & 0.65 \\
$S_{1v}^{\dagger} \overline{O} S_{2v} + cc $ & 2.18 & 0.62  & 2.27 & 2.75 \\
$S_{2v}^{\dagger} \overline{O} S_{2v} + cc $ & 10.11 & 2.56 & 14.26 & 11.75 \\
\hline\\ 
\multicolumn{5}{c}{\textbf{Important effective two-body terms of $\overline{O}$ }} \\
\hline
 & & & & \\
$S_{2v}^{\dagger} O T_1 + cc $ & -0.72 & -0.05 & -0.02 & -0.02 \\
$S_{2v}^{\dagger} O T_2 + cc $ & -5.19 & -0.76 & 0.43 & 0.46 \\
 $Norm$ & -4.89 & -0.93 & -2.02 & -2.07 \\
\hline
\hline
\end{tabular}
\label{tab:front4}
\end{table}

\section{conclusion}
The RCC theory has been employed to study the magnetic dipole and electric quadrupole hyperfine 
structure constants of the $^{205}$Pb$^+$. Strong electron correlations effects are found in 
the $d-$ states and their behavior is different from other systems studied
earlier. Experiments to measure these quantities will constitute important 
tests of the relativistic coupled-cluster theory. 

\section{Acknowledgment}
We are grateful to Prof. Werth for valuable discussions and suggestions
for this calculation. The calculations were carried out using the Tera-flop
Supercomputer at C-DAC, Bangalore, India.

\end{document}